\documentclass[12pt]{article}

\usepackage[paper=a4paper, margin=2cm]{geometry}
\usepackage[russian,english]{babel}
\usepackage[cp1251]{inputenc}
\usepackage{amsmath}

\usepackage[logos]{amsbib}

\begin{document}

\date{}
\renewcommand{\refname}{References}

\author{A.~Yu.~Samarin
}
\title{Can quantum objects be point-like particles?}

\maketitle

\centerline{Samara Technical State University, 443100 Samara, Russia}
\centerline{}

\abstract{\textsl{The possible nature of the nonlinear evolution of quantum systems is explored from the point of view of nonlocal effects under the conditions of EPR paradox. It is shown that both stochastic and deterministic mechanisms of nonlocal evolution should consider a quantum particle as a material field in order not to contradict special relativity. It is assumed that there is an experiment allowing to find out which of the mechanisms of nonlinear evolution is realized in nature.} 



The existence of the nonlinear evolution of the state of an open system combined with a non-local procedure of the wave function normalization generates a seeming possibility of a nonlocal interaction between the remote parts of a composite quantum system in entangled state~\cite{bib:1}. In general, such a possibility violates basic relativistic constraints. The inadmissibility of this violation imposes restrictions on possible physical interpretations of mathematical objects and procedures used in quantum mechanics. 

There are two types of these restrictions. The first one is based on the fact that the nonlinear evolution equation allows us to keep these interactions local in the case of statistical mixtures. Then, if to introduce a nonlinear stochastic term in the equation, the nonlocal correlations of the probabilities taking place in the experiments like~\cite{bib:2,bib:3,bib:4} (performed by the scheme~\cite{bib:5}) could be substantiated without contradiction with special relativity~\cite{bib:6,bib:7,bib:8}. This approach generates two undesirable consequences. The first one is that the insertion of stochastic terms in fundamental equations attaches  probabilities nonepistemic character\footnote{Such a probability character in the conventional interpretation of quantum mechanics~\cite{bib:9} is in question: the observables are the result of the measuring process taking place with the participation of a macroscopic apparatus. Thus, between the probabilities and the fundamental deterministic Schr\"odinger equation there is another process the nature of which is not known reliably.}(It is unlikely that the solution of one of the many problems that have arisen in physics requires such a radical change in the attitude to the principle of sufficient cause.). The second one is that possible states of the particles of a single pair after measurement are in a deterministic correspondence between themselves regardless of the choice of the measured quantity \footnote{If there are no hidden parameters and we assume a quantum particle in the form of a point-like particle, then, this fact limits the scope of special relativity to a macroscopic level.}.

In addition to the described method of formal "...peaceful coexistence between quantum mechanics and relativity"~\cite{bib:6}, there is a possibility that the violation of the relativistic requirements is farfetched and there is no problem in principle. Really, for all the varieties of the EPR paradox, it is assumed that events in entangled subsystems are space-like separated, although this does not follow from the formalism of quantum mechanics. Let us consider Einstein's statement--- "Every element of the physical reality must have a counterpart in the physical theory"---~\cite{bib:10} and assume that the converse statement is also true. Then every term of the expansion for stationary states of a wave function has a physical reality. Therefore, if we follow the realistic approach that any physical quantity (including a wave function) is the attribute of a carrying agent, we can conclude that a quantum particle is a collection of matter fields (continuums). These continuums occupy all space accessible\footnote{The accessible space is determined by the previous history of the quantum particle and the environment.} for the particle even there, where the wave function is zero. Thus, there is no space-like interval between the measurement on the particle $A$ and the reduction of the particle $B$ state (we use the conventional designations) under the conditions of the EPR paradox\footnote{The apparatuses are only remote from each other, quantum particles are material fields in the same region of space.}.

At the present time, it seems that any other possible interpretations of the quantum theory that are compatible with special relativity contradict the results of the EPR experiments carried out according to the Bell scheme. Thus, we have a choice between the stochastic evolution of ensembles of point-like quantum particles and the deterministic evolution of material fields\footnote{It will be shown below that the stochastic evolution mechanism does not allow to conserve the image of a quantum particle in the form of a point-like object.}. This alternative has a different physical basis, which, in principle, allows for experimental verification.

Let us consider two conventional observers $A$ and $B$ which can operate with a remote apparatus measuring the quantities of the entangled composite system. Denote by $A$ and $B$ the corresponding parts (subsystems) of the system. Let the composite system be closed. Denote by $q$ the set of the subsystem $A$ coordinates, by $x$, $y$ the coordinates sets of the subsystem $B$. For eigenvalue expansion of the wave function of a closed system with respect to the measurable quantity of the subsystem A, we have
\begin{equation*}
 \Psi(x,q,t_{0})=\sum\limits_j a_j(x,t_{0})\varphi_j(q)
\end{equation*}
The reduced density matrix of the subsystem $B$ before the measurement on the system $A$ in the form of expansion on the eigenstates of the subsystem $A$ is 
\begin{eqnarray*}
 \rho(y,x,t_{0})=\int\Psi^*(y,q,t_{0})\Psi(x,q,t_{0})\,dq=\int\sum\limits_{i}a_i^*(y,t_{0})\varphi_i^*(q)\sum\limits_j a_j(x,t_{0})\varphi_j(q)\,dq=\\
 =\sum\limits_{i}a_j^*(y,t_0)a_j(x,t_0),
\end{eqnarray*}
Expending this density matrix in the series on eigenstates of an observable of the subsystem $B$ (that will be measured), we obtain
\begin{equation*}
 \rho(y,x,t_{0})=\sum\limits_i\sum\limits_{m}\sum\limits_{n}b_{im}^*(t_0)b_{in}(t_0)\phi_m^*(y)\phi_n(x)
\end{equation*}
Thus, for the elements of the reduced density matrix of the subsystem $B$ (in the representation of the measured quantity) before the measurements, we have   
\begin{equation}
 w_{mn}=\sum\limits_iw^i_{mn},
\end{equation}
where
\begin{equation*}
w^i_{mn}= b_{i,m}^*(t_0)b_{i,n}(t_0).
\end{equation*}
If, after the measurement on the subsystem $A$, we have the result $k$, then, the reduced density matrix of the subsystem $B$ takes the form (for the approach 2):
\begin{equation}\label{eq:math:ex2}
 w_{mn}=\sum\limits_i\frac{w^i_{mn}\delta_{ik}}{trw^i} =\frac{w^k_{mn}}{trw^k},
\end{equation}
where $\delta_{ik}$ is the Kronecker symbol. Such a situation takes place for the deterministic evolution of the state of a single particle B. If we have  the statistical ensemble of the particles B states and do not know which result has been obtained (denote by $p_k$ the probabilities of the measured results), then, for the reduced density matrix we obtain 
\begin{equation}\label{eq:math:ex3}
 w_{mn}=\sum\limits_kp_k\frac{w^k_{mn}}{trw^k} =\sum_k w^k_{mn}.
\end{equation}
Expressions~\eqref{eq:math:ex2},~\eqref{eq:math:ex3} are identical for both the stochastic and deterministic\footnote{in this case the probabilities are generated due to the statistical straggling of the macroscopic apparatus performing the successive measurement.} approach to the reduction process, but, for consistency with special relativity, both require the nonlocality of quantum objects in the case if the ensemble $B$ is not a mixture (a homogeneous ensemble, for example). For the first time, it was proposed to create such an ensemble by cloning the state of a single photon~\cite{bib:11}. But according to the no-cloning theorem, this turned out to be impossible~\cite{bib:12}. Nevertheless, such an ensemble could result from the simultaneous reduction of the set of particles $B$ under the measurement performed on the subsystem $A$, which is in an entangled state with each of the particles $B$~\cite{bib:13}. The consequence of the realization of such an experiment is the evidence of the fact of the quantum objects nonlocality. But it can not answer the question about the stochastic nature of the collapse. The following experiment can do this.

Let the subsystems $A$ and $B$ be macroscopic. Let the evolution of the coherent state corresponding to one of the system's ($A+B$) degrees of freedom be described by a wave function\footnote{For example the wave function of an electronic state of the atoms ensemble in the coherent electromagnetic field.}. If the subsystem $A$ is affected by an external coherent influence\footnote{Such as the radiation of a quantum oscillator.} nonlinearly transforming its state, then, in accordance with the deterministic evolution law~\cite{bib:1}, the state of the subsystem $B$ transforms simultaneously. Since both the system and the subsystems are macroscopic, these changes are not random and can be detected definitely\footnote{This, of course, means the realization of the so-called a faster-then-light communication.}. If a nonlinear evolution is random, this effect does not exist.

\vfill\eject


\begin{thebibliography}{10}

\Bibitem{bib:1}
\by A. Yu. Samarin
\paper Nonlinear dynamics of open quantum systems
\arxiv \href{http://arxiv.org/abs/1706.09405}{1706.09405} [quant-ph]
\yr 2017

\Bibitem{bib:2}
\by J.F. Clauser, M.A. Horne, A. Shimony and R.A. Holt
\paper Proposed Experiment to Test Local Hidden-Variable Theories
\jour Phys. Rev. Lett.
\vol 23
\pages 880--883
\yr 1969
\crossref{http://dx.doi.org/10.1103/PhysRevLett.23.880}

\Bibitem{bib:3}
\by S.J. Freedman and J.F. Clauser
\paper Experimental Test of Local Hidden-Variable Theories
\jour Phys. Rev. Lett.
\vol 28
\pages 938--941
\yr 1972
\crossref{http://dx.doi.org/10.1103/PhysRevLett.28.938}


\Bibitem{bib:4}
\by A. Aspect
\paper Bell's inequality test: more ideal than ever
\jour Nature
\vol 398
\pages 189--190
\yr 1999
\crossref{http://dx.doi.org/10.1038/18296}

\Bibitem{bib:5}
\by J. S. Bell
\paper On the Einstein-Podolsky-Rosen paradox
\jour Physics
\vol 1
\pages 195--200
\yr 1964


\bibitem{bib:6}
\paper Stochastic quantum dynamics and relativity
\by Gisin~N.
\jour :	Helvetica physica acta
\vol 62
\pages 363--371
\yr 1989
\url{http://cms.unige.ch/gap/quantum/wiki/_media/publications:bib:stochqdynrel.pdf}

\Bibitem{bib:7}
\paper Correlation experiments in nonlinear quantum mechanics
\by Czachor~M.,Doebner~H\,D.
\jour Phys. Lett. A
\vol 301
\pages 139--152
\yr 2002
\crossref{http://dx.doi.org/10.1016/S0375-9601(02)00959-3}
\arxiv \href{http://arxiv.org/abs/quant-ph/0110008v2} [quant-ph]

\Bibitem{bib:8}
\paper Dynamical reduction models
\by Bassi~A.,Ghirardi~G\,C.
\jour Phys.Rept.
\vol 379
\pages 257--426
\yr 2003
\crossref{http://dx.doi.org/10.1016/S0370-1573(03)00103-0}
\arxiv \href{http://arxiv.org/abs/quant-ph/0302164v2} [quant-ph]

\Bibitem{bib:9}
\book  Mathematical foundations of quantum mechanics
\by von Neumann J.
\publ Princeton University Press
\yr 1955
\publaddr Princeton

\Bibitem{bib:10}
\by A. Einstein, B. Podolsky and N. Rosen
\paper Can quantum mechanics description be considered complete?
\jour Phys. Rev.
\vol 47
\pages 777--780
\yr 1935
\crossref{http://link.aps.org/doi/10.1103/PhysRev.47.777}

\Bibitem{bib:11}
\paper FLASH"---A superluminal communicator based upon a new kind of quantum measurement
\by Herbert~N.
\jour :	Foundations of Physics
\vol 12
\pages 1171--1179
\yr 1982
\url{http://link.springer.com/article/10.1007%2FBF00729622#page-1http://link.springer.com/article/10.1007%2FBF00729622#page-1}

\Bibitem{bib:12}
\paper A single quantum cannot be cloned
\by Wootters~W.~K., Zurek~W.~H. 
\jour :	Nature 
\vol 299
\pages 802--803
\yr 1982
\url{http://www.nature.com/nature/journal/v299/n5886/abs/299802a0.html}
\url{https://arxiv.org/ftp/quant-ph/papers/0210/0210060.pdf}

\Bibitem{bib:13}
\by ~Samarin~A.\,Yu.
\paper Nonlocal transformation of the internal quantum particle structure 
\jour J. Samara State Tech. Univ. Ser. Phys. and Math. Sci. 
\yr 2016
\vol 20(3)
\pages 423--456
\mathnet{http://mathnet.ru/vsgtu1484}



\end{thebibliography}
\end{document}